\documentclass[a4paper,10pt]{article}

\usepackage[utf8]{inputenc}
\usepackage{palatino}
\usepackage{amsmath}
\usepackage{graphicx}
\usepackage[italian,english]{babel}
\usepackage{SIunits}
\usepackage[margin=2cm]{geometry}
\usepackage{url}
\usepackage{multicol}

\let\at=@
\catcode`@=\active
\def@#1{\boldsymbol{#1}}

\begin{document}
\selectlanguage{italian}
\title{Fisica e fumetti: Paperone ed il deposito sotterraneo}
\author{Franco Bagnoli$^{1,2}$ e Francesco Saverio Cataliotti$^{1,3}$\\[1ex]
  \begin{minipage}{\textwidth}
  \small
    \begin{enumerate}
    \item Dip. Energetica, Università di Firenze and INFN, sez.\ Firenze, via S. Marta 3 50139 Firenze, Italia
    \item  Also CSDC, Univ.\ Firenze email: franco.bagnoli\at unifi.it
    \item  Also LENS, Univ.\ Firenze email: francescosaverio.cataliotti\at unifi.it
    \end{enumerate}
  \end{minipage}
}

\maketitle
\begin{abstract}
I fumetti, come i film, spesso utilizzano idee scientifiche ``fantasiose''. 
Non ci riferiamo qui alla \emph{violazione
implicita} delle leggi della fisica, cosa permessa in un mondo di
fantasia, quanto piuttosto all'uso di \emph{spiegazioni} fisiche errate
che vengono usate in buona fede perché a riflettono convinzioni molto
diffuse, ma sbagliate, sull'interpretazione di fenomeni a partire
dai principi fisici. D'altra parte questi errori possono
servire a illustrare la corretta applicazione della fisica in una
maniera molto più accattivante rispetto alla modalità tradizionale
di presentazione. Analizziamo qui l'avventura \emph{Paperone ed il
deposito sotterraneo} di Pezzin e Cavezzano~\cite{Paperone}.
\end{abstract}

\selectlanguage{english}
\begin{abstract}
Comics and cartoon movies sometimes exploit fictitious scientific ideas. 
It is often the case that these ideas, althought wrong, actually reflect the popular vision of some natural phenomenon. We do not refer here to the \emph{implicit violation} of physical laws in fictions, a practice allowed by the underlining ``poetic licence'' of comics.  However, sometimes wrong scientific ``explanations'' are proposed, and those may be accepted by the public without further inspection. 
On the other hand, these errors may be a good starting point for a didactic illustration of physical principles. We analyze here the comics \emph{Paperone ed il
deposito sotterraneo} (Uncle Scrooge and the underground money bin) by Perrin e Cavezzano~\cite{Paperone}.
\end{abstract}

\selectlanguage{italian}

\begin{multicols}{2}
\section{Introduzione}
\begin{figure*}
  \begin{center}
    \begin{tabular}{cc}
      \includegraphics[width=0.25\textwidth]{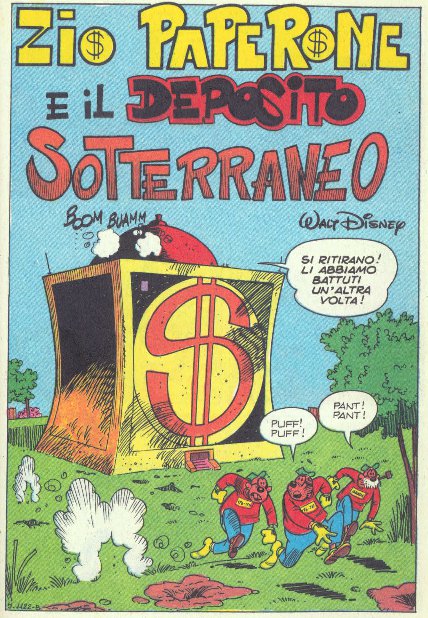}  & \includegraphics[width=0.25\textwidth]{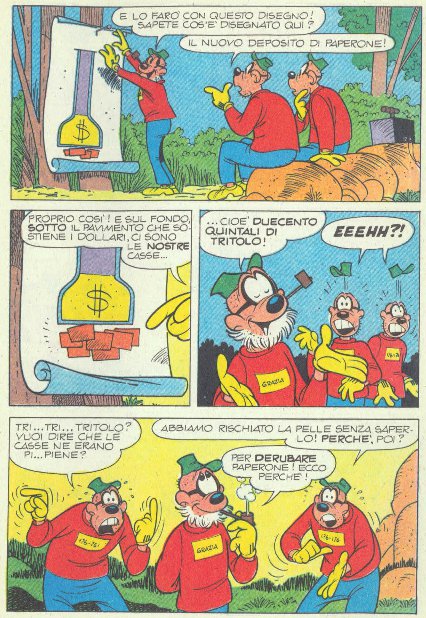}\\
      (a) & (b)
    \end{tabular}
  \end{center}
  \caption{\label{a} (a) Prima pagina dell'avventura \emph{Paperone ed il deposito sotterraneo}. (b) Lo schema del deposito sotterraneo e la carica nascosta dai Bassotti.}
\end{figure*}

\begin{figure*}
  \begin{center}
    \begin{tabular}{ccc}
      \includegraphics[width=0.25\textwidth]{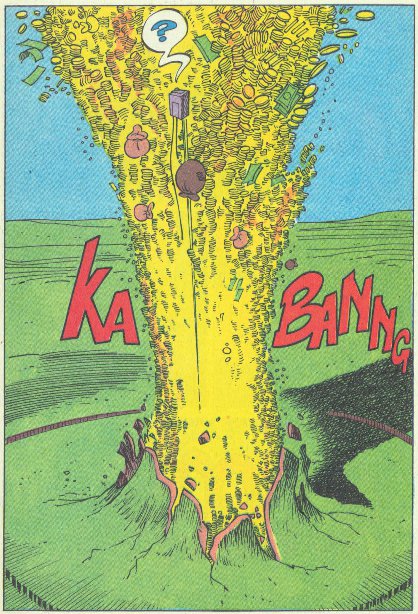}  & \includegraphics[width=0.25\textwidth]{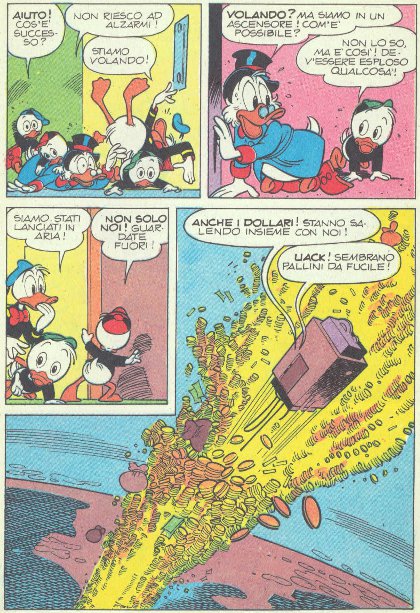} & \includegraphics[width=0.25\textwidth]{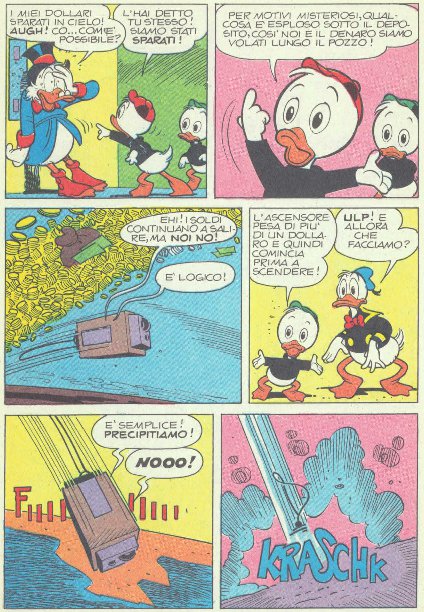} \\
      (a) & (b) &(c)
    \end{tabular}
  \end{center}
 \caption{\label{b} (a) Esplosione del deposito. (b) Traiettoria di dollari ed ascensore. (c) Separazione tra ascensore e dollari.}
\end{figure*}

L'avventura \emph{Paperone ed il deposito
sotterraneo}~\cite{Paperone} inizia con Paperone esasperato dai
continui tentativi di furto da parte della Banda Bassotti
(fig.~\ref{a}-a). Per cercare di scoraggiare altri tentativi di
furto, Paperone decide di costruire un deposito sotterraneo.
Purtroppo, i Bassotti vengono a sapere del piano e durante la
costruzione, mascherati da operai, riescono a nascondere delle casse
di dinamite sotto il pavimento del deposito sotterraneo
(fig.~\ref{a}-b). Mentre Paperone, Paperino e i nipoti Qui, Quo e
Qua stanno scendendo con l'ascensore, i Bassotti fanno esplodere la
dinamite, cosicché dollari e ascensore (contenente i paperi)
vengono scagliati in alto (fig.~\ref{b}-a). I paperi sopravvivono
alla tremenda accelerazione; dopo poco però si accorgono che
l'ascensore si separa dalla massa dei dollari e inizia a scendere (fig.~\ref{b}-b).
Uno dei paperini trova subito la spiegazione: \emph{E' logico!
L'ascensore pesa di più di un dollaro e quindi comincia prima a
scendere!} (fig.~\ref{b}-c).

\begin{figure*}
  \begin{center}
    \begin{tabular}{ccc}
      \includegraphics[width=0.25\textwidth]{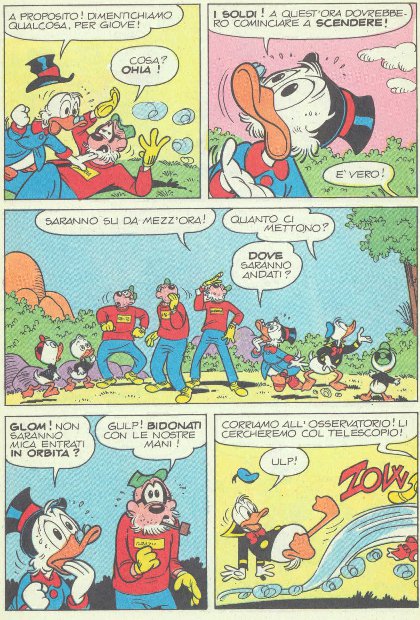}  & \includegraphics[width=0.25\textwidth]{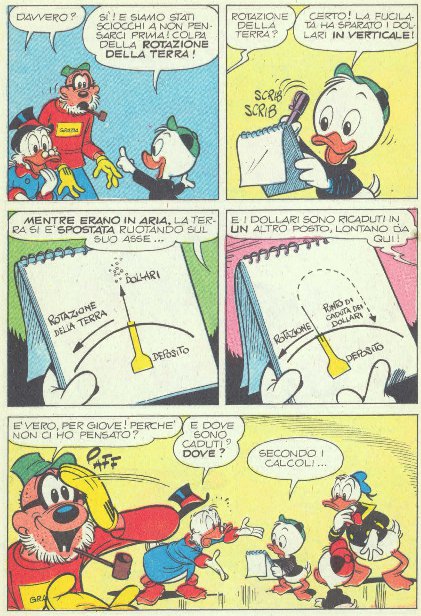} & \includegraphics[width=0.25\textwidth]{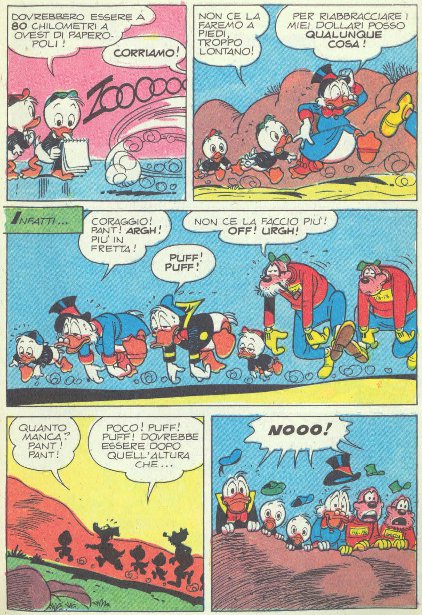}\\
      (a) & (b) & (c)
    \end{tabular}
  \end{center}
 \caption{\label{c} (a) La scomparsa dei dollari. (b) La ``spiegazione''. (c) La distanza stimata.}
\end{figure*}

Dopo essere sopravvissuti anche all'atterraggio, i
paperi ed i Bassotti attendono invano il rientro dei dollari, e
temono che possano essere entrati in orbita (fig.~\ref{c}-a). Anche in questo caso,
è uno dei paperini che ``scopre'' il motivo della scomparsa dei
dollari: \emph{Siamo stati sciocchi a non pensarci prima! Colpa della
rotazione della Terra!} [\ldots ] \emph{Mentre erano in aria la Terra si è
spostata ruotando sul suo asse!} (fig.~\ref{c}-b) [\ldots ] \emph{Secondo i calcoli dovrebbero
essere \emph{[atterrati]} a $80~\kilo\metre$ a ovest di Paperopoli!}
(fig.~\ref{c}-b). Ovviamente la conclusione non è così semplice.
Dopo aver percorso gli $80~\kilo\metre$ di corsa, scopriranno che i
dollari sono finiti in mare e Paperone costringerà i Bassotti a
recuperarli.

La storia presenta un certo numero di errori di fisica. Non ci
riferiamo qui alla capacità di sopravvivere ai traumi, cosa che
appartiene alla ``fisica dei fumetti'', che possono essere
assimilati ai cartoni animati. Come spiegato esplicitamente in
Ref.~\cite{Roger}, i cartoni possono morire solo dal ridere o se
vengono sciolti nella ``salamoia'', mentre si riprendono dai traumi
senza conseguenze, riportando al limite un leggero stordimento. Ovvero: \emph{L'animazione segue le leggi della fisica, a meno che il contrario risulti più divertente.}~\cite{Babbit}.  Si veda anche l'appendice~\ref{cartoni} sulle leggi del moto nei cartoni animati. 

Per un autore di fumetti la fisica (e tutte le scienze) costituiscono stimoli immaginifici che non necessariamente 
si traducono in una trattazione accurata. G. Pezzin, autore della storia, commenta~\cite{pezzin}
\begin{quotation}
 \it
 In effetti non avevo fatto veri calcoli scrivendo la storia. A me interessava di più l'effetto
complessivo. L'idea l'ho avuta leggendo appunto Verne, ma soprattutto mi piaceva l'immagine di una ``fucilata di dollari'', soprattutto se disegnata da Cavazzano.
\end{quotation}
Tuttavia, quando si invocano esplicitamente le leggi della fisica
\emph{umana}, data la popolarità dei fumetti si corre il rischio di 
lasciare \emph{tracce} pericolose nei futuri studenti di fisica, ingegneria, ma non solo\dots Cerchiamo qui di \emph{rimediare}, per quanto possibile.

Gli errori che si
possono rimarcare sono:
\begin{enumerate}
 \item Durante il volo in caduta libera, dopo l'esplosione, i paperi stanno ritti sul pavimento come se questo fosse in quiete.
 \item L'ascensore si separa dalla massa dei dollari in virtù del suo peso.
 \item Si teme che i dollari siano entrati in orbita.
 \item Il punto di  caduta si sposta verso ovest di $80~\kilo\metre$ a causa della rotazione terrestre.
\end{enumerate}

Dopo aver introdotto le varie forze agenti sugli oggetti del
problema e aver considerato gli ordini di grandezza coinvolti,
discuteremo i primi tre errori nella sezione~\ref{sec:inizio},
mentre dedicheremo la sezione~\ref{sec:rotazione} all'analisi
dell'ultimo errore nel caso di moto nel vuoto, e la sezione~\ref{sec:attrito} al caso con attrito. Discuteremo i possibili ulteriori approfondimenti e il valore didattico dell'esempio nella sezione
finale.

\section{Qualche commento sulle forze coinvolte e ordini di grandezza}

Nel seguito eseguiremo alcuni calcoli sulla base di modelli, come
per esempio quello del punto materiale. Ogni modello nasce in base a
certe approssimazioni della realtà, e quindi è necessario
controllare gli ordini di grandezza dei vari fattori. E' spesso
inutile eseguire dei calcoli complessi sulla base di un modello che
non è giustificato sulla base di quanto si conosce del sistema
sperimentale. D'altra parte, a volte un modello troppo semplice, pur
permettendo di eseguire facilmente dei calcoli, può non essere
adeguato.

Eseguiremo i calcoli nel sistema di riferimento accelerato
dell'osservatore sulla superficie terrestre. L'accelerazione di
gravità $g$ in funzione della distanza dal centro della terra $r$
è $g=G M_\odot/R^2 _\odot$, 
dove $M_\odot  \simeq  5.98 \cdot   10^{24}~\kilogram$ è la massa terrestre,
$R_\odot  \simeq  6.372 \cdot  10^{6}~\metre$ è il raggio medio terrestre e
$G \simeq  6.67 \cdot 
10^{-11}~\metre\cubed\per\kilogram\per\second\squared$ la costante
di gravitazione universale. Scrivendo $r=R_\odot + h$, con $h$ altezza dal suolo, si ha
\[
g= \frac{G M_\odot }{(R_\odot+h)^2}\simeq \frac{G M_\odot }{R^2 _\odot }\left(1-2 \frac{h}{R_\odot }\right).
\]
Quindi l'approssimazione di forza peso indipendente dalla quota ha
un errore del 1\% circa ogni $31.5~\kilo\metre$ di altezza raggiunta.

Detta $v_0 $ la velocità iniziale, il tempo di volo in assenza di attrito dell'aria è $\tau  = 2v_0 /g$.
L'altezza $h_m$ raggiunta (in assenza di attrito dell'aria) è $h=v^2 _0 /(2g)$.

In presenza di attrito dell'aria ($\boldsymbol F_a=-\gamma  \boldsymbol v$), con coefficiente
$\gamma$ in effetti dipendente dalle dimensioni e dalla forma
dell'oggetto in moto, si ha
\[
h=\frac{m v_0 }{\gamma }-\frac{m^2  g}{\gamma ^2 }
\log\left(1+\frac{\gamma  v_0 }{m g}\right).
\]

L'altezza di $31.5~\kilo\metre$ si raggiunge in assenza di attrito
dell'aria per una velocità iniziale di circa $790
~\metre\per\second$. In presenza di attrito dell'aria con un
coefficiente realistico $\gamma =0.1~\kilogram\per\second$, la velocità~\kilogram\per\second
necessaria per una massa $m=1~\kilogram$ è circa $3500~\metre\per\second$ (molto superiore alla
velocità dei proiettili). D'altra parte il moto non   è
sicuramente viscoso (ma piuttosto turbolento), e inoltre l'atmosfera
è spessa solo qualche chilometro, con grandi variazioni di densità
e temperatura, questo porterebbe ad utilizzare una formula
differente per l'attrito dell'aria con un coefficiente dipendente
dall'altezza.

Vediamo dunque le forze  che agiscono su di un proiettile:
\[
 \begin{split}
  \boldsymbol F_{peso} &= - m g \boldsymbol j, \\
  \boldsymbol F_{aria}&= -\gamma  \boldsymbol v, \\
  \boldsymbol F_{Coriolis}&= -2 m \boldsymbol {\omega } \wedge  \boldsymbol v, \\
  \boldsymbol F_{Centrifuga}&= -m  \boldsymbol {\omega }\wedge (\boldsymbol {\omega }\wedge  \boldsymbol r).
\end{split}
\]

Per un corpo di massa 1 kg assumendo $g=9.81~\metre\per\second\squared$, la velocità di rotazione terrestre
$\omega \simeq 7.27\cdot 10^{-5}~\rad\per\second$ e $\gamma =0.1~\kilogram\per\second$, si hanno  i rapporti fra i moduli (con $v_0 =1000~\metre\per\second$)

\[
 \begin{split}
  \frac{F_{aria}}{F_{peso}}&= 10.2, \\
  \frac{F_{Coriolis}}{F_{peso}}&= 0.0148, \\
  \frac{F_{Centrifuga}}{F_{peso}}&=0.0034.
\end{split}
\]

Ne segue che non si può trascurare la forza di attrito
dell'aria, pur non sapendo bene come modellizzarla. La forza centrifuga è trascurabile perché piccola
e diretta come la forza peso (all'equatore) mentre la forza di
Coriolis NON è trascurabile perché, pur piccola, è
diretta perpendicolarmente alla forza peso (all'equatore).

Eseguiremo  comunque i calcoli prima in assenza di attrito da parte
dell'aria, e vedremo che si possono ottenere i risultati con carta e
penna. Passeremo poi alla discussione in presenza di attrito
viscoso, ma solo come esercizio di calcolo. Approssimeremo
$g=10~\metre\per\second\squared$.

\section{Moto dei gravi}\label{sec:inizio}

Il primo errore di fisica consiste nel fatto di rappresentare i
paperi in piedi nell'ascensore anche se questo è in caduta libera.

\begin{figure*}
  \begin{center}
    \begin{tabular}{ccc}
      \includegraphics[height=0.22\textheight]{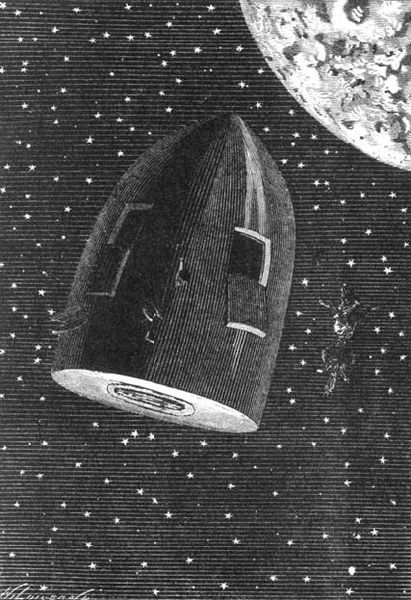}  &
      \includegraphics[height=0.22\textheight]{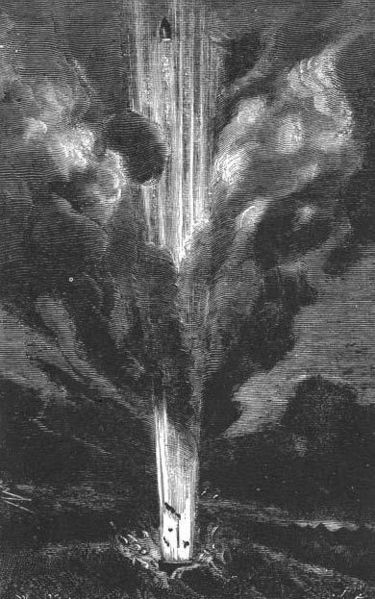}  &
      \includegraphics[height=0.22\textheight]{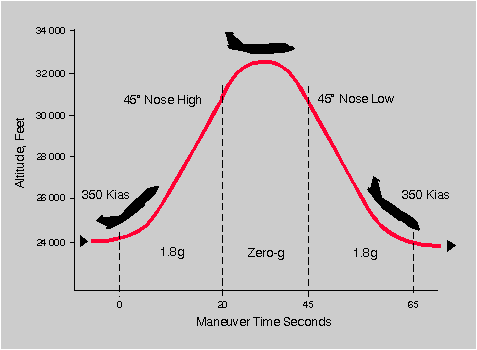}  \\
      (a)&(b)&(c)
    \end{tabular}
  \end{center}
 \caption{\label{proiettile} (a) Il proiettile ed il cadavere del cane in una illustrazione dal libro \emph{Intorno alla Luna}~\cite{verne2} disegnata da Émile-Antoine Bayard e Alphonse de Neuville (16 Sep. 1872). (b) Il lancio del proiettile in un disegno dall'edizione illustrata di \emph{Viaggio dalla Terra alla Luna}~\cite{verne} del 1872. (c) La traiettoria dell'aereo (soprannominato ``vomit comet'') utilizzato per le riprese e gli esperimenti in assenza di gravità.}
\end{figure*}

Facciamo prima le considerazioni nel vuoto. 
Il sistema di riferimento dell'ascensore è un sistema in moto
accelerato con accelerazione pari a $g$. All'interno di un tale
sistema di riferimento si osserva una forza fittizia di valore $-m
g$ cioè tale da cancellare esattamente la forza peso. Quindi,
relativamente all'ascensore, i paperini dovrebbero galleggiare
``senza peso''; In effetti gli esperimenti (e i film) in condizioni di
gravit\`{a} ridotta vengono proprio realizzati in sistemi in caduta
libera, per esempio all'interno di un aereo in stallo~\cite{esa} (Fig.~\ref{proiettile}-c) o in appositi
contenitori lasciati cadere da una torre, senza dimenticare le navicelle spaziali in orbita intono alla terra (che probabilmente nel prossimo futuro diventeranno convenienti economicamente anche per la cinematografia) o la stazione spaziale 
internazionale. 

Si può anche fare riferimento al principio di equivalenza debole, ovvero al fatto che localmente non è possibile distinguere un campo gravitazionale dall'accelerazione di un sistema di riferimento non inerziale, principio che facilmente porta a calcolare la deviazione di un raggio luminoso in un campo gravitazionale.

Senza ricorrere ai sistemi di riferimento non-inerziali, quando un
sistema è soggetto solo alla forza di gravità, e questa si può
considerare omogenea, nessun sottosistema viene accelerato
diversamente dal resto. L'errore nel fumetto può servire a mettere
in evidenza che noi, con i nostri sensi, non ``sentiamo'' la forza
di gravità, che agisce nella stessa maniera in ogni nostra parte.
Quello che ci dà il senso del ``peso'' è in realtà la reazione
vincolare del suolo, che ci impedisce di sprofondare, trasmessa
attraverso le nostre ossa al resto del corpo. La stessa spiegazione
fisica è alla base dei traumi che vengono causati dalle brusche
accelerazioni (o più probabilmente. dalle decelerazioni, per
esempio durante un incidente automobilistico). Quello che ci causa
danni non è la decelerazione, ma il fatto che venga impartita solo
ad una parte del nostro corpo da un vincolo (per esempio il
parabrezza della nostra auto).

Rimanendo all'interno della letteratura di fantasia, che le brusche
accelerazioni causino danni è, per esempio, ben presente in J.
Verne, che, nel romanzo ``Dalla Terra alla Luna''~\cite{verne}, a
cui in fondo questo fumetto si ispira, usa degli ammortizzatori ad
acqua per permettere ai protagonisti di sopravvivere (anche se uno
 dei due cani imbarcati morirà per i traumi subiti). La massima accelerazione sopportabile senza danni da un essere umano varia con la posizione del corpo, per una posizione ``generica'' non deve superare i $3g$ ($g$ è l'accelerazione di gravità, $9.8~\metre\per\second$), ma può arrivare a più di $10g$ se il corpo è sdraiato. Durante un urto può superare le centinaia di $g$. 

Viceversa, lo
stesso Verne nel romanzo ``Intorno alla Luna''~\cite{verne1}
commette un errore simile a quello qui esaminato quando descrive
come l'effetto di ``assenza di peso'' si possa apprezzare solo per
un breve momento vicino al punto di equilibrio tra le attrazioni
terrestre e lunare (vedere l'appendice~\ref{verne}).

Non è forse inutile sottolineare come il brindisi finale non
avrebbe potuto aver luogo -- i liquidi in caduta libera possono
rimanere compatti a causa della tensione superficiale, ma certo non
possono essere versati nei bicchieri. D'altra parte, Jules Verne
giustamente sottolinea che il cadavere del cane morto, gettato poi
nello spazio, segue il proiettile per tutto il viaggio
(fig.~\ref{proiettile}-a).

E' possibile anche proporre un facile esperimento di verifica~\cite{referee}, utilizzando una bottiglia di plastica (senza tappo) riempita d'acqua in cui è stato praticato un forellino vicino al fondo. Lasciandola cadere da qualche metro di altezza si vede come il flusso di acqua cessi immediatamente appena inizia la caduta, e che l'acqua non esce neppure dal foro del tappo. 
Con un po' di esercizio, si riesce a lanciare la bottiglia in verticale senza farla ruotare, ed anche in questo caso il flusso cessa al momento del lancio.

Tornando ai nostri paperi, dato che il moto dell'ascensore e dei
dollari avviene in aria, si dovrebbe tenere conto dell'attrito
viscoso, che può influenzare anche la discussione dell'errore n. 3.
In caso di moto viscoso, si dovrebbe avere come risultato che
l'ascensore viene rallentato rispetto alla caduta libera e quindi i
paperi dovrebbero (forse) potersi tenere in piedi. Dalla figura si
può però desumere che i dollari e l'ascensore viaggiano in
formazione compatta, trascinando presumibilmente con sé anche
l'aria. In questo caso ovviamente non ha senso la separazione
dell'ascensore dai dollari, in quanto le forze agenti (la forza di
gravità $mg$) è proporzionale alla massa così come la forza di
inerzia ($ma$). Dall'equivalenza tra massa inerziale
e massa gravitazionale, ogni oggetto in caduta libera si muove con
la stessa accelerazione indipendentemente dalla sua massa. In
effetti solo l'attrito dell'aria, introducendo un forza non
linearmente dipendente dalla massa è in grado di giustificare
una separazione fra ascensore e dollari, ritorneremo su questo punto
nel prossimo paragrafo~\ref{sec:attrito}.

Il terzo errore è un po' più tecnico. In principio è possibile
sparare un proiettile con una velocità tale da metterlo in orbita o
addirittura farlo uscire dalla gravità terrestre~\cite{verne}.
Anzi, questo avrebbe il grande vantaggio di non dover accelerare anche il combustibile, per esempio il carico utile del razzo Saturn V (il razzo che ha portato gli uomini sulla
luna) è molto basso: 3000 tonnellate totali  per 118 tonnellate di carico utile in
orbita bassa (47 tonnellate per la luna)~\cite{saturn}.

La
prima velocità cosmica $v_1$ è la velocità che bisogna imprimere
ad un proiettile o missile per permettergli di entrare in orbita
circolare con raggio minimo. Dall'equilibrio tra forza
gravitazionale e forza centrifuga
\[
  m \dfrac{v_1 ^2 }{r} = G \dfrac{M_\odot m}{r^2 },
\]
in cui $M_\odot \simeq 5.98\cdot  10^{24}~\kilogram$ è la massa terrestre,
$m$ la massa del proiettile, $r$ è il raggio dell'orbita, $G\simeq 
6.67 \cdot  10^{-11}~\metre\cubed\per\kilogram\per\second\squared$ la
costante di gravitazione universale, si ottiene, prendendo
$r=R_\odot \simeq  6.372 \cdot  10^6  \metre$, raggio terrestre per un'orbita
radente alla superficie, (appena al di sopra dell'atmosfera),
\[
  v_1 \simeq  8 \cdot  10^3 ~\metre\per \second.
\]

D'altra parte, la quantità di polvere da sparo (e la lunghezza
della canna) necessaria per imprimere una certa velocità ad un
proiettile cresce notevolmente con la velocità del proiettile
stesso. Citando dalla Ref.~\cite{earmi},

\begin{quotation}
\it
  Se per 1000 m/s la carica di polvere pesava il 40\% del peso del proiettile, per ottenere 1300 m/s occorreva un peso di polvere pari a quello del proiettile.
\end{quotation}

Il progetto HARP~\cite{harp} mirava appunto a usare cannoni come
lanciatori per oggetto in alta atmosfera, ma non raggiunse mai
velocità superiori a meno della metà della velocità $v_1$.
Tuttavia, ci sono nuovi progetti del genere in atto~\cite{sharp}.

Dalla fig.~\ref{a} è evidente che la carica non è abnorme rispetto al contenuto del deposito. Del resto, possiamo desumere la velocità iniziale $v_0 $ dalla stima (errata) fatta dai paperini (Fig.~\ref{c}-c): la velocità tangenziale della terra è circa $437~\metre\per\second$, e per uno spostamento di $80~\kilo\metre$ si ha un tempo di volo di $\tau  \simeq  183~\second$. Dato che $\tau  = 2v_0 /g$ si ottiene $v_0 \simeq 915~\metre\per\second$, che noi approssimiamo a $1000~\metre \per \second$, ben al di sotto della prima velocità cosmica.

La variazione della forza di gravità con l'altezza è trascurabile, per cui nel seguito useremo $g$ costante (e uguale a $10~\metre\per\second\squared$). Approssimeremo coerentemente $\tau  \simeq  200~\second$.

\section{Rotazione della terra e sistemi di riferimento non inerziali}
\label{sec:rotazione}

Il quarto errore merita una discussione più ampia. La
``spiegazione'' del paperino è ovviamente sbagliata. I dollari,
essendo sparati in verticale, mantengono la stessa velocità
tangenziale della terra. Se il movimento potesse essere assimilato
ad un moto traslatorio uniforme, i dollari dovrebbero ricadere nello
stesso punto. Ma il moto della terra non è traslatorio, così che
ci troviamo in un sistema di riferimento accelerato. Prendiamo
quindi un sistema di riferimento locale in cui l'asse $y$ sia
verticale (con lo zero sulla superficie terrestre), l'asse $x$ vada
da est a ovest  e l'asse $z$  sia perpendicolare a questi due (terra
vista dal polo nord). Facciamo innanzi tutto i calcoli come se il
deposito si trovasse sull'equatore. La rotazione della terra avviene
in senso  antiorario con una velocità angolare  $\omega = 2\pi /(24 \cdot 
3600)~\rad\per \second \simeq  7.27 \cdot  10^{-5}~\rad\per\second$ e per
il momento la consideriamo diretta come l'asse $z$, ovvero $\boldsymbol {\omega }
=\omega  \boldsymbol k$.

Le forze ``apparenti'' in un sistema di riferimento non inerziale
sono la forza centrifuga  $- m\boldsymbol {\omega } \wedge  (\boldsymbol {\omega } \wedge  \boldsymbol r)$ e la forza
di Coriolis $-2 m\boldsymbol {\omega } \wedge  \boldsymbol v$.  Come osservato in precedenza, si
pu\`{o} trascurare il contributo della forza centrifuga. Rimane
quindi il contributo della forza di Coriolis. Impostando le
equazioni del moto si ottiene
\begin{align}
   m \ddot x   &= 2 m \omega  \dot y ,            \label{ddotx}\\
   m \ddot y  &= - m g - 2 m \omega  \dot x .     \label{ddoty}
\end{align}

Integrando l'eq.~\eqref{ddotx} tra 0 e $t$, con le condizioni
iniziali $y(0)=0$ ed $\dot{x}(0)=0$ cioè dollari sparati in
verticale, si ottiene
\begin{equation}
   \dot x  = 2 \omega  y ,                 \label{dotx}
\end{equation}
che è sempre positivo. Quindi effettivamente il proiettile viene
spostato dalla rotazione terrestre verso ovest, ma per una ragione
diversa da quella ipotizzata. E' probabilmente più facile
visualizzare l'effetto in un sistema di riferimento fisso (osservatore inerziale): al
momento dello sparo i dollari, oltre alla velocità verticale, hanno
una componente orizzontale (tangenziale alla terra) diretta da ovest
verso est. Il moto è quasi parabolico, data limitata altezza
raggiunta, ma al momento in cui i dollari raggiungono la retta
tangente alla terra e passante dal loro punto di lancio, sulla
verticale (quasi) del loro punto di partenza, devono ancora
``scendere'' per un tratto dato che la terra è sferica. Durante
questa caduta residua la terra continua a girare e quindi i dollari
cadono ad ovest.

Calcoliamo precisamente traiettoria e punto di caduta nelle nostre
approssimazioni.
Sostituendo l'eq.~\ref{dotx} nella \ref{ddoty} si ottiene
\begin{equation}
 \ddot y  = -  g - 4  \omega ^2  y,              \label{osc}
\end{equation}
ovvero l'equazione di un oscillatore armonico, cosa che può sorprendere un po'. La soluzione con le condizioni iniziali $y(0)=0$, $\dot y (0) = v_0 $ è
\begin{equation}
 y(t) = \dfrac{g}{4  \omega ^2 }\left(\cos(2\omega t)-1\right) + \dfrac{v_0 }{2\omega }\sin(2\omega t).              \label{y}
\end{equation}
Nel limite $\omega \rightarrow 0$ si ottiene, sviluppando %il seno al primo ordine e il coseno
al secondo ordine,
\[
 y(t) \simeq  v_0  t - \dfrac{1}{2} g t^2 ,
\]
come ci aspettavamo.

\begin{figure*}
  \begin{center}
    \begin{tabular}{cc}
      \includegraphics[height=0.22\textheight]{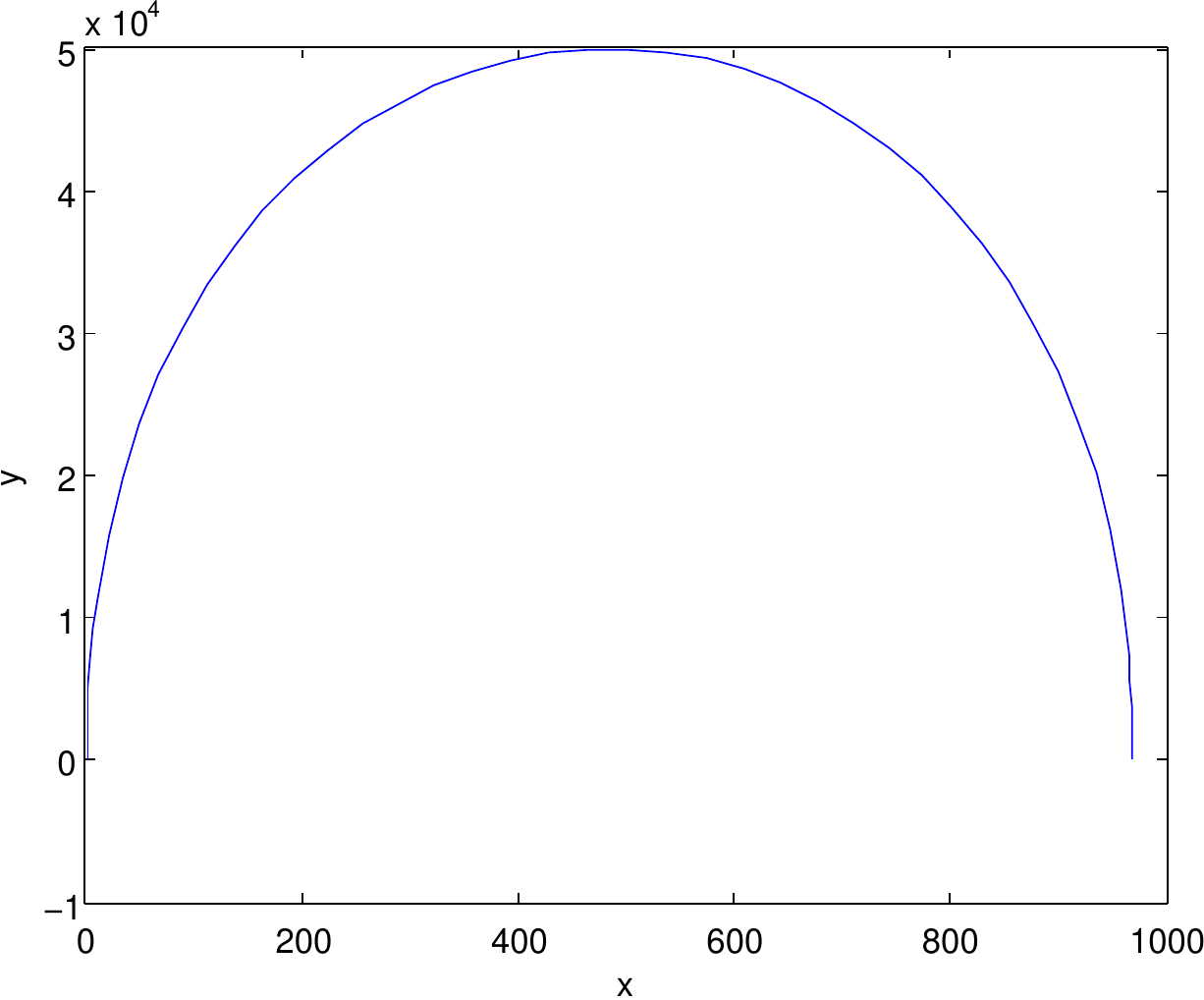}  &
      \includegraphics[height=0.22\textheight]{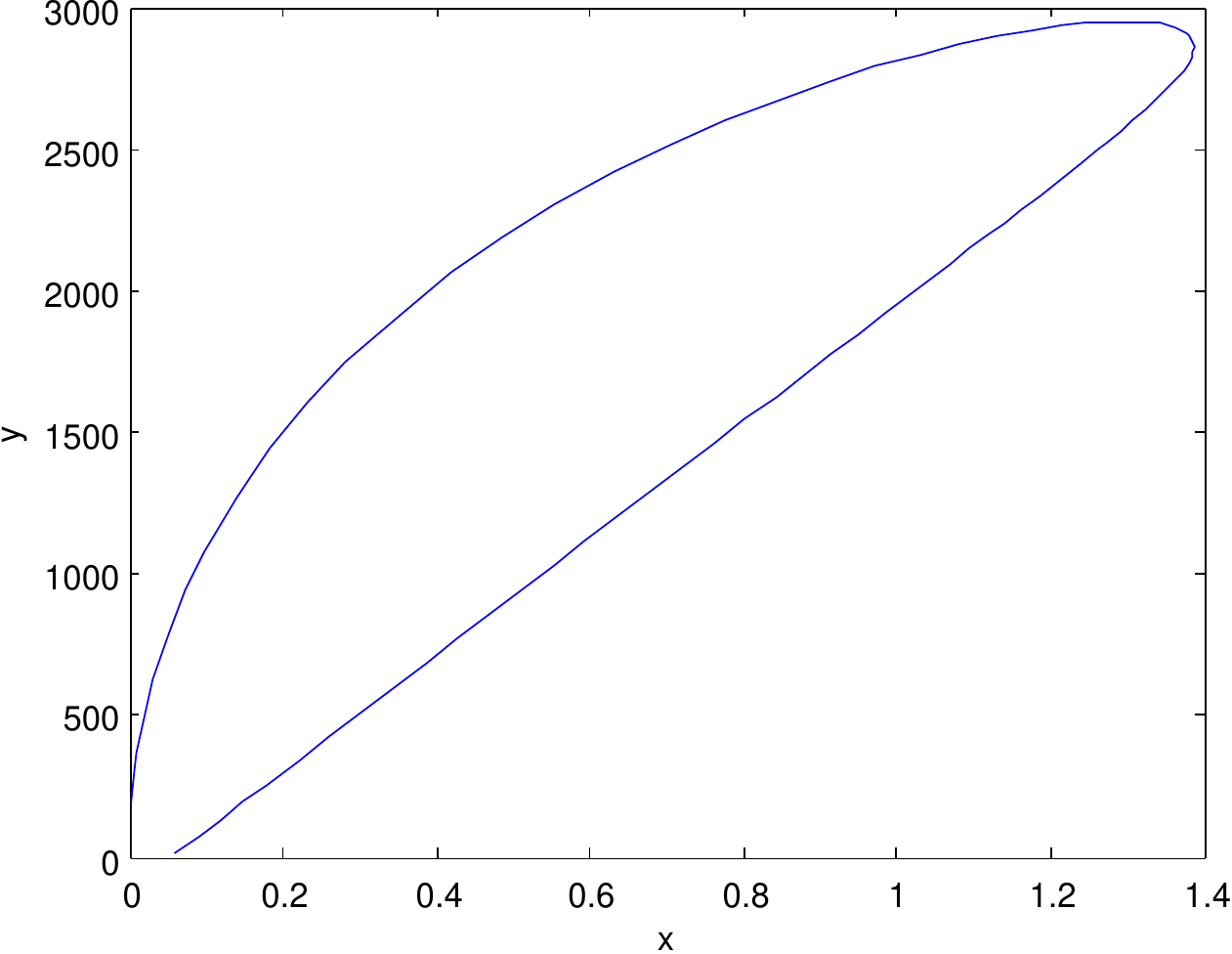}   \\
      (a)&(b)
    \end{tabular}
  \end{center}
 \caption{\label{traiettoria} (a) Traiettoria dell'ascensore in assenza (linea tratteggiata) e in presenza (linea continua) di aria, $\omega = 2\pi /(24 \cdot 3600)~\rad\per \second \simeq  7.27 \cdot  10^{-5}~\rad\per\second$,  $g=10~\metre\per\second\squared$, $\gamma=0.1~\kilogram\per\second$, $m=1000~\kilogram$. (b) Traiettoria dell'ascensore in presenza di attrito viscoso dell'aria, assumendo $m=0.03~\kilogram$.}
\end{figure*}

Sostituendo l'eq.~\ref{y} nell'eq.~\ref{dotx} ed integrando con le condizioni iniziali $x(0)= \dot x (0)=0$ si ottiene
\begin{equation}
 x(t) = \dfrac{g}{4 \omega ^2 }\left[\sin(2\omega t) - 2\omega t\right] - \dfrac{v_0 }{2\omega }\left[\cos(2\omega t)-1\right].  \label{x}
\end{equation}
Possiamo verificare che nel limite $\omega \rightarrow 0$ si ottiene al primo ordine
$x(t)=0$. Sviluppando al terzo ordine otteniamo
correttamente
\begin{equation}
x(t)\simeq\omega\left(v_0t^2-\frac13gt^3\right)
\end{equation}

Al tempo $\tau $ si ha
\[
 x(\tau ) \simeq  \dfrac{4\omega v_0 ^3 }{3g^2 }  =1~\kilo\metre
\]
(fig.~\ref{traiettoria}-a), ovvero una distanza che si può agevolmente percorrere a corsa
(anche se senza allenamento probabilmente si arriva senza fiato),
come illustrato in fig.~\ref{c}.

Paperopoli però non è sull'equatore. Dato che si trova in
California si può pensare che sia più o meno a $37\degree $N. In questo
caso si avrà una deviazione verso ovest di soli $765~\metre$.

%\subsection{Metodo alternativo}
%
%La maniera più semplice per ottenere questo risultato è quella di
%integrare l'equazione~\eqref{dotx} tra 0 e $\tau $ usando per $y$
%l'equazione del moto in assenza di rotazione $y =  v_0  t - (1/2)
%gt^2$.

Si noti che la procedura a prima vista corretta per cui si assume che un proiettile
rimanga "indietro" rispetto al punto di sparo della distanza
$d=\omega  \int_0 ^\tau y(t)dt$, partendo dal'assunto che la velocità tangenziale
alla superficie sia minore di quella in quota, porta ad un risultato che è la metà di quello corretto (dato il fattore 2 nella forza di Coriolis).

\section{Calcolo in presenza di attrito viscoso}\label{sec:attrito}

Per esercizio possiamo sviluppare il calcolo in presenza di attrito viscoso, proporzionale alla velocità. Dal secondo principio della dinamica

\begin{align}
\label{smorzx}
  \ddot x  &= 2\omega \dot y -(\gamma /m) \dot x , \\
  \ddot y  &=-g - 2 \omega  \dot x  - (\gamma /m) \dot y .\label{ddotyg}
\end{align}
Ricavando $\dot y $ e $\ddot y $ dalla prima di queste equazioni e
sostituendo nella seconda otteniamo

\begin{equation}
  \dddot{x}+(\gamma /m)\ddot x =-2\omega  g-4\omega ^2 \dot x -(\gamma /m)\ddot x -(\gamma /m)^2 \dot x ,
\end{equation}
con le sostituzioni

\[
  \begin{split}
    \Gamma  &= \gamma /m, \\
    \Omega  &= \sqrt {4\omega ^2 +(\gamma /m)^2 },\\
    \zeta  &= \dot x +2\omega  g/\Omega ^2 ,
  \end{split}
\]
si ottiene l'equazione di un moto armonico smorzato

\begin{equation}
  \ddot{\zeta}+2\Gamma \dot{\zeta}+\Omega ^2 \zeta=0.
\end{equation}

Risolvendo questa equazione con le condizioni iniziali
$x(0)=0$, $\dot x (0)=0$, $\ddot x (0)=2\omega  v_0 $, si ottiene

\[
\begin{split}
 x=
  \frac{1}{\Omega ^4}\Bigl\{&
    2\omega (2g\Gamma +\Omega ^2  v_0 )\\
     &-e^{-\Gamma t}\bigl[
      2\omega (2g\Gamma +\Omega ^2 v_0 )\cos(2\omega  t))\\
      &+\left(g\Gamma ^2 +\Gamma \Omega ^2 v_0 -4g\omega ^2 \right)\sin(2\omega t)\bigr]
      -2g\omega \Omega ^2 t\Bigr\}.
\end{split}
\]

Inserendo la soluzione nell'equazione per $\dot y $ ed integrando si
ottiene

\[
\begin{split}
y=\frac{1}{\Omega ^4}\Bigl\{&\left(g\Gamma ^2 +\Gamma \Omega ^2 v_0 -4g\omega ^2 \right)\\
   &+ e^{-\Gamma  t}\bigl[2\omega (2g\Gamma +\Omega ^2  v_0 )\sin(2\omega  t)\\
    &-\left(g\Gamma ^2 +\Gamma \Omega ^2  v_0 -4 g\omega ^2 \right)\cos(2\omega t)\bigr]
    -g\Gamma \Omega ^2 t\Bigr\}.
\end{split}
\]

All'ordine zero in $\omega $ dà la soluzione del moto in assenza di
forza di Coriolis, che si discosta poco dalla soluzione reale.

\begin{equation}
y=\frac{1}{\Gamma }\left[\left(v_0 +\frac{g}{\Gamma }\right)\left(1-e^{-\Gamma t}\right)-g t\right].
\end{equation}

Per corpi di  massa piccola rispetto al coefficiente di attrito  ($\Gamma =\gamma/m$ grande),
la forza di attrito viscoso domina sulla forza
di Coriolis anche nel caso del moto lungo l'asse $x$;
l'accelerazione lungo l'asse $x$ sar\`{a} perci\`{o} nulla.
Integrando l'equazione~\eqref{smorzx} con le condizioni iniziali
$x(0)=0$ ed $y(0)=0$ possiamo quindi ottenere
\begin{equation}
x(t)\simeq-\frac{2 m\omega}{\gamma}y(t),
\end{equation}
e quindi al termine del moto i dollari ricadono quasi esattamente
sul punto di partenza. 
La traiettoria numerica dei dollari (assumendo una massa di $30~\gram$) è riportata in Fig.~\ref{traiettoria}-b.
Si noti che nel caso di moto viscoso la traiettoria è quasi verticale: la scala orizzontale nella figura~\ref{traiettoria}-b è di soli $1.4~\centi\metre$. Inoltre l'altezza massima è molto ridotta, ed il tempo di volo (non indicato) si riduce a circa la metà ($\tau\simeq 100~\second$).  Ovviamente si è trascurato il trascinamento dell'aria da parte della massa dei dollari, per cui probabilmente si dovrebbe osservare un notevole sparpagliamento delle monete.

Per l'ascensore, invece, il valore di $\Gamma$ è molto piccolo, data la grande massa. Prendendo sempre $\gamma=0.1~\kilogram\per\second$ e assumendo $m=1000~\kilogram$, la traiettoria dell'ascensore non si discosta essenzialmente da quella in assenza di aria (spostamento finale circa $950~\metre$ all'equatore). Le traiettorie dell'ascensore calcolate numericamente in assenza e presenza di aria sono riportate in Fig.~\ref{traiettoria}-a.

Quindi, in presenza di aria, è presumibile che lo scenario sia opposto di quello ipotizzato nel fumetto: l'ascensore devia verso ovest molto più dei dollari.

\section{Conclusioni}

Abbiamo presentato un'analisi della fisica di un fumetto cercando di mettere in luce gli aspetti più semplci da calcolare anche con carta e penna. 
E' ovviamente possibile svolgere il calcolo in maniera 
accurata, anche se ciò comporta l'uso dell'integrazione numerica. Riteniamo comunque che anche in maniera
approssimata la discussione presenti degli spunti interessanti.

Ovviamente non intendiamo contestare qui la ``licenza poetica'' da parte degli autori di opere di fantasia  
che può giustificare qualsiasi violazione delle leggi della fisica. 
D'altra parte, la presentazione di leggi fisiche a partire da un fumetto dovrebbe
stimolare la curiosità e la riflessione, soprattutto se presentata
come ``esercizio di scoperta'', tipo: ``quali errori di fisica sono
stati commessi in questo fumetto?''.

\section*{Ringraziamenti}
Ringraziamo Giorgio Pezzin per la piacevole conversazione e i consigli e Gianluca Martelloni per i prezioni suggerimenti e la verifica dei calcoli analitici. 

\appendix

\section{Le leggi del moto nei cartoni animati} \label{cartoni}
di Mark O'Donnell~\cite{leggi}\\

\textbf{Un corpo sospeso nello spazio non cade finché non diventa conscio della propria situazione.}
\textit{Paperino corre fuori da un dirupo, credendo che ci sia ancora la terra sotto ai suoi piedi. Vaga nell'aria chiacchierando tra sé finché decide di guardare in basso. A questo punto, la nota legge della caduta dei gravi riprende a valere.}

\textbf{Un corpo in movimento permane nel suo stato di moto finché un altro corpo solido non interviene all'improvviso.}
\textit{I personaggi dei cartoni animati, che siano sparati da un cannone o stiano rincorrendo qualcuno, hanno una tale quantità di moto che solo un palo del telegrafo o un enorme macigno possono fermarli.}

\textbf{Un corpo che passa attraverso un muro lascia un buco della sua stessa identica forma.}
\textit{Questo fenomeno, noto anche come "sagoma da attraversamento", è tipico di chi subisce la spinta di un'esplosione, o dei personaggi codardi, così ansiosi di scappare da uscire direttamente dal muro, lasciando la propria forma perfettamente ritagliata. L'arrivo di un farabutto (o di un matrimonio) spesso catalizza questa reazione.}

\textbf{Il tempo impiegato da un oggetto a cadere per 20 piani è maggiore o uguale al tempo impiegato da chiunque lo abbia fatto cadere dal davanzale a fare 20 spirali nel vuoto nel tentativo di riprenderlo intatto.}
\textit{Tale oggetto è inevitabilmente inestimabile; il tentativo di riprenderlo, inevitabilmente vano.}

\textbf{La legge di gravità può essere violata dalla paura.}
\textit{La forza della mente è sufficiente in molti corpi a provocare una spinta che li porti lontano dalla superficie terrestre. Un rumore sinistro o il verso tipico del nemico provocheranno un moto verso l'alto, di solito verso un lampadario, il ramo di un albero o la cima dell'asta di una bandiera. I piedi di uno che corre o le ruote di un'auto in velocità possono anche non toccare mai terra, portando chi fugge a volare.}

\textbf{Al crescere della velocità, gli oggetti possono trovarsi in più posti contemporaneamente.}
\textit{Ciò accade particolarmente nei combattimenti corpo a corpo, in cui si può intravedere la testa di un personaggio fuoriuscire dalla nube di una lite in più luoghi simultaneamente. L'effetto è molto comune anche nei corpi che roteano vorticosamente, e stimola il nostro modo di vedere trattenendo le immagini. Alcuni personaggi riescono ad auto-replicarsi solamente a velocità folli, e possono continuare a rimbalzare contro i muri prima di raggiungere la velocità richiesta per questo fenomeno di auto-replicazione della propria massa.}

\textbf{Alcuni corpi possono attraversare un muro su cui è disegnato l'ingresso di una galleria; altri no.}
\textit{Questa contraddizione del trompe l'oeil lascia sconcertati da generazioni, eppure è noto che chiunque disegni un ingresso finto sulla superficie di un muro per raggirare un nemico, non sarà mai in grado di inseguirlo in questo spazio virtuale. Colui che ha disegnato si appiattisce regolarmente contro il muro ogni volta che ci prova. Questo però è un problema per critici d'arte, non per scienziati.}

\textbf{La necessità, sommata alla volontà, può causare materializzazioni spontanee.}
\textit{Quando la paura di un pericolo improvviso lo rende necessario, oggetti incredibilmente solidi - come martelli, candelotti di dinamite, torte, seducenti abiti da donna - possono spuntare all'improvviso da dove prima si credeva ci fosse solo lievissima aria. Alcuni spiegano questo controverso fenomeno della "tasca senza fondo" pensando che questi oggetti spuntino da invisibili recessi dei vestiti dei protagonisti, o provengano da un magazzino appena fuori dallo schermo; ma questo sposta la questione su come si possa ogni volta trovare istantaneamente l'oggetto desiderato.}

\textbf{Ogni disintegrazione violenta di materia felina è temporanea.}
\textit{I gatti dei cartoni animati hanno molte più vite delle usuali nove. Possono essere affettati, appiattiti, piegati a fisarmonica, arrotolati o fatti a pezzi, ma non vengono distrutti. Dopo qualche istante di bieca autocommiserazione, essi si rigonfiano, riaccorciano, rimontano e solidificano.}

\textbf{Ad ogni vendetta corrisponde un'altra vendetta uguale e contraria.}
\textit{Questa è l'unica legge dei cartoni animati che si applica molto bene anche al mondo reale. Proprio per questo ci piace vedere che accada ad un papero, piuttosto che a noi.}

\textbf{Chiunque sia in caduta libera raggiunge il terreno più velocemente di un'incudine.}
\textit{Tale incudine cadrà immancabilmente sulla sua testa.}

\textbf{Un corpo appuntito tende a spingere improvvisamente un personaggio verso l'alto.}
\textit{Quando un personaggio viene punto (per esempio nel fondoschiena) da un tale corpo (per esempio uno spillo), infrange la legge di gravità lanciandosi verso l'alto a folle velocità.}

\textbf{Le armi esplosive non possono causare danni permanenti.}
\textit{Esse hanno l'unico effetto di rendere i personaggi temporaneamente neri e bruciacchiati.}

\textbf{La gravità si trasmette tramite onde lunghe a bassa velocità.}
\textit{Si può osservare questo fatto quando un personaggio si trova sospeso improvvisamente su un baratro. Prima cominciano a scendere i piedi, causando un allungamento delle gambe. Quando l'onda gravitazionale raggiunge il torso, esso comincerà a cadere causando l'allungarsi del collo. Infine, quando anche la testa sarà stata raggiunta dall'onda, la tensione si rilasserà e il personaggio riprenderà le proporzioni usuali, finché non raggiungerà il suolo.}

\textbf{La dinamite si genera spontaneamente negli "spazi C" (gli spazi in cui valgono le leggi dei cartoni animati).}
\textit{Il processo è analogo alla teoria dello stato stazionario dell'universo, che ipotizza che le forze che tengono insieme l'universo causino la creazione dal nulla di atomi di idrogeno. I quanti di dinamite sono piuttosto grandi e instabili. Tali quanti sono attirati dalle forze psichiche generate dalla sensazione di pericolo di alcuni personaggi, che riescono a usare la dinamite a proprio vantaggio. Si può pensare che gli "spazi C" siano il risultato di una primordiale esplosione di dinamite (un big-bang, in effetti).}

\section{Dalla Terra alla Luna}\label{verne}

 [ \ldots ] Difatti la traiettoria del proiettile correva dalla Terra alla Luna. A
mano a mano che si allontanava dalla Terra, l'attrazione terrestre
diminuiva in ragione inversa dei quadrati delle distanze, ma anche
l'attrazione lunare aumentava nella stessa proporzione. Doveva,
dunque, arrivare un punto dove le due attrazioni si neutralizzavano e
quindi il proiettile avrebbe perso tutto il suo peso. Se le masse della
Luna e della Terra fossero state eguali, questo punto sarebbe stato
equidistante dai due astri. Ma, data la differenza delle masse, era
facile calcolare che il punto era situato a quarantasette
cinquantaduesimi del viaggio e cioè, in cifre, a settantottomila cento
e quattordici leghe dalla Terra.

In questo punto un corpo, che non avesse avuto in sé nessun
principio di velocità o di spostamento, sarebbe rimasto eternamente
immobile, perché attirato con forza eguale dai due astri, senza che
nulla lo sollecitasse più verso l'uno che verso l'altro.[\ldots ]

Ora, come riconoscere se il proiettile aveva raggiunto
il punto neutro a settantottomila cento e quattordici leghe dalla
Terra?

Lo avrebbero riconosciuto al preciso momento in cui né essi né gli
oggetti del proiettile, avrebbero più subito la legge di gravità.
Fin qui i viaggiatori, anche percependo che quell'azione andava
diminuendo sempre di più, non ne avevano riconosciuto ancora
l'assenza totale. Ma quel giorno, verso le undici del mattino, Nicholl
si fece scappare un bicchiere di mano e il bicchiere, invece di cadere,
rimase sospeso nell'aria.

— Oh! — esclamò Michel — ecco un'esperienza fisica davvero
divertente.

Subito diversi oggetti, armi, bottiglie abbandonate a se stesse
rimasero librate come per miracolo. Anche Diana, posta nello spazio
da Michel, riprodusse, ma senza nessun trucco, la meravigliosa
sospensione fatta dai Caston e dai Robert-Houdin. Né la cagnetta
sembrava accorgersi d'esser librata nell'aria.

Essi stessi, sorpresi, stupefatti, nonostante i loro ragionamenti
scientifici, sentivano che in quel dominio del meraviglioso nel quale
erano arrivati, da uomini avventurosi, il peso mancava ai loro corpi.

Se stendevano le braccia, esse non tendevano più ad abbassarsi. Le
teste vacillavano sulle spalle. I piedi non aderivano più al fondo del
proiettile. Erano come ubriachi che non hanno più stabilità. La
fantasia ha creato uomini privi di riflessi, altri privi di ombra! Ma qui
la realtà, in virtù della neutralità delle forze attrattive, formava
uomini nei quali più nulla pesava e che non pesavano più nulla essi
stessi.[\ldots ]

I suoi amici lo raggiunsero in un istante e tutti e tre, al centro del
proiettile, rappresentavano una miracolosa ascensione.

— Ma è una cosa credibile? È verosimile? È possibile? —
esclamava Michel Ardati. — No. Eppure è vero! Ah! se Raffaello ci
avesse visti così, che Assunzione avrebbe buttato giù sulla sua tela!

— L'Assunzione non durerà — rispose Barbicane. — Se il
proiettile passa il punto neutro, l'attrazione lunare ci attirerà verso la
Luna.

— E allora i nostri piedi poggeranno sulla volta del proiettile —
rispose Michel.

— No — disse Barbicane — perché il proiettile, che ha il centro
di gravità molto in basso, piano piano si rivolterà.

— Ma allora tutta la nostra sistemazione sarà messa sotto sopra, è
la parola giusta!

— Rassicurati, Michel — rispose Nicholl. — Non c'è da temere
nessun caos. Gli oggetti non si muoveranno poiché l'evoluzione del
proiettile avverrà insensibilmente.

— Infatti — interloquì Barbicane — e quando esso avrà lasciato il
punto di attrazione eguale, la sua culatta, relativamente più pesante,
lo trascinerà seguendo un movimento perpendicolare alla Luna. Ma
perché questo fenomeno si produca, bisognerà che oltrepassiamo la
linea neutra.

— Oltrepassare la linea neutra! — esclamò Michel. — Allora
faremo come i marinai che tagliano l'equatore. Brindiamo al nostro
passaggio!

Un leggero movimento di fianco riportò Michel verso la parete
imbottita. Là egli prese una bottiglia e i bicchieri, li mise «nello
spazio», dinanzi ai suoi amici e, bevendo allegramente, tutti e tre
salutarono la linea con un triplice urrà!

Quell'influenza delle attrazioni durò un'ora appena.~\cite{verne2}

\end{multicols}
\end{document}